%Paper: hep-th/9207047
%From: app@cuphyf.phys.columbia.edu (Alexios Polychronakos)
%Date: Tue, 14 Jul 92 11:14:14 EDT

%%%%%%%%%%%%%%%%%%%%%%%%%% BODY OF PAPER %%%%%%%%%%%%%%%%%%%%%%%%%%%%%%

\input phyzzx
\Pubnum={CU-TP-569\cr
UB-ECM-PF-92/16\cr}
\date={June 1992}
\titlepage
\title{Schr\"odinger Equation for Particle with Friction}
\bigskip
\author{Alexios~P.~Polychronakos\footnote\dag
{app@cuphyf.phys.columbia.edu; address after November 1992:
CERN, Theory Division, CH-1211 Gen\`eve 23, SWITZERLAND.}}
\address{Pupin Physics Laboratories, Columbia University,\break
New York, NY 10027, USA}
\andauthor {Rodanthy Tzani\footnote\ddag
{tzani@ebubecm1.bitnet}}
\address{Departament d' Estructura i Constituents de la Mat\'eria,\break
Universitat de Barcelona, E-08028 Barcelona, SPAIN}
\bigskip
\abstract{
A new quantum mechanical wave equation describing a particle
with frictional forces is derived. It depends on a parameter
$\alpha$ whose range is determined by the coefficient of friction
$\gamma$, that is, $0 \leq \alpha \leq \gamma$. For one extreme value of
this parameter, $\alpha = 0$, we recover Kostin's equation. For the other
extreme value, $\alpha = \gamma$, we obtain an equation in which friction
manifests in ``magnetic" type terms. It further exhibits breakdown of
translational invariance, manifesting through a symmetry breaking parameter
$\beta$, as well as localized stationary states in the absence of external
potentials. Other physical properties of this new class of equations are
also discussed.}
\endpage

\def\AP{{\it Ann. Phys.\ }}
\def\PTP{{\it Prog. Theor. Phys.\ }}
\def\NP{{\it Nucl. Phys.\ }}

\def\PR{{\it Phys. Rev. \ }}
\def\PRD{{\it Phys. Rev. D\ }}
\def\PRA{{\it Phys. Rev. A\ }}
\def\PRL{{\it Phys. Rev. Lett.\ }}

\def\JCP{{\it J. Chem. Phys.\ }}

\def\ZP{{\it Z. Phys.\ }}
\REF\bw{V.~Bargmann and E.P.~Wigner, {\it Proc. Nat. Acad. Sci. U.S.A.}
{\bf 34} (1948) 211.}
\REF\jn{V.P.~Nair and R.~Jackiw, \PRD {\bf 43} (1991) 1993.}
\REF\caletal{C.G.~Callan and L.~Thorlacius, \NP {\bf B329} (1990) 117;
C.G.~Callan, A.G.~Felce and D.E.~Freed, PUPT-1292.}
\REF\nak{S.~Nakamiya, \PTP {\bf 20} (1958) 948.}
\REF\sen{I.R.~Senitzky, \PR {\bf 150} (1960) 670.}
\REF\zw{R.~Zwanzig, \JCP {\bf 33} (1960) 1338.}
\REF\mor{H.~Mori, \PTP {\bf 33} (1965) 423.}
\REF\caleg{A.O.~Caldeira and A.J.~Legget, {\it Physica} {\bf A121}
(1983) 587; \PRL {\bf 46} (1981) 211; \AP {\bf 149} (1983) 374.}
\REF\kos{M.D.~Kostin, \JCP {\bf 57} (1972) 3589.}
\REF\dek{H.~Dekker, \ZP {\bf B21} (1975) 295; \PRA {\bf 16} (1977) 2116.}
\REF\ya{K.~Yasue, \AP {\bf 114} (1978) 479.}
\REF\nel{E.~Nelson, \PR {\bf 150} (1966) 1079.}

%Definitions
\def\half{{1\over2}}
\def\a{\alpha}
\def\b{\beta}
\def\g{\gamma}
\def\d{{\rm d}}
\def\D{{\rm D}}
\def\p{\psi}
\def\P{\psi^*}
\def\f{\Phi}
\def\F{\Phi^*}
\def\r{\rho}
\def\t{\theta}
\def\e{\epsilon}
\def\ve{\varepsilon}

It is known that not all classical dynamical systems can be successfully
quantized. Apart from problems such as nonrenormalizability, topological
obstructions, anomalies etc., there is a whole class of systems for
which there is no available method of quantization, namely
non-hamiltonian systems. The most common examples are systems
with dissipation, although these are by no means the only ones. The
Bargmann-Wigner higher spin field equations [\bw] and field theories of
fundamental anyons [\jn] are notable cases too.
The ``canonical" example of a non-hamiltonian system is, nevertheless,
the case of a particle on the line with friction. Its equation of motion is
$$
m \ddot x + \g \dot x + {dV \over dx} = 0
\eqn\fric$$
where overdot denotes time derivative and $V(x)$ is the potential
of the particle. There is no local action functional which produces
\fric\ as the equation of motion, and therefore no canonical structure
and no quantum mechanics. This model (and its higher dimensional
versions), apart from its interest as a phenomenological description of
dissipation due to interactions, also has recently found applications in
string theory [\caletal].

There are two points of view one could adopt on the previous problem.
One is to consider it as a fundamental physical question as to whether
and how such systems can be quantized. The other is to consider such
dissipative systems as phenomenological descriptions of more fundamental
(hamiltonian) systems interacting with a many degrees of freedom reservoir,
in which the dissipation is a macroscopic manifestation of these
interactions. The question, then, is to find an adequate quantum mechanical
description of these systems without explicitly involving the degrees
of freedom of the reservoir. The first question is more fundamental
and mathematically intriguing, while the second is more physical.

There have been several attempts to this problem, which can be classified
into two similar categories as above: those who start with an initial
expanded system including the reservoir and then ``integrate out" the extra
degrees of freedom [\nak-\caleg], and those who start on the outset with a
modified quantization scheme which classically reduces to the desired
dynamics [\kos,\dek]. The first approach is clearly more physically
motivated. It has, however, some disadvantages. For instance, coupling a
particle to a string of oscillators to reproduce dissipation and then
integrating them out, leads to a state where the fourth moment of the
particle position diverges, that is
$$
< x^{2n} > = \infty \,\,\,\, {\rm for}\,\,\,\, n \geq 2
\eqn\div$$
This is because the zero-point motion of the infinite set of
oscillators perturbs the particle in a substantial way.
Obviously this is an unsatisfactory feature which would hopefully
be absent in a fundamental quantization procedure of dissipative
systems. It is also not clear that the answers obtained are in general
model-independent.

Using the second approach, Kostin [\kos] has proposed a modified
Schr\"odinger equation for the model \fric\ in which the friction
is reproduced through a wavefunction-dependent potential. This
has some unusual and rather controversial properties: it violates the
superposition principle, and has stationary states in which the
energy does not dissipate. (This equation was rederived by Yasue [\ya]
using Nelson's stochastic quantization scheme [\nel].) Although the above
properties are not necessarily fatal (we {\it know} that the
quantum mechanics of these systems must be radically different
from the standard one), probably because of them Kostin's equation
has not been used very extensively.

The purpose of this paper is to present an alternative Schr\"odinger
equation for this process. In fact, we will derive a family of
equations which contains Kostin's as a special case. Physical properties
of these new equations, in comparison with Kostin's, will also be
discussed.

The main new feature of our approach consists of allowing a
wavefunction-dependent {\it first}-derivative term in Schr\"odinger's
equation, that is
$$
i \dot \p = -\half \d^2 \p + W \d \p + ( U+V ) \p
\eqn\prenew$$
where $\psi$ is the wavefunction and $W$, $U$ are potentials
explicitly depending on $\psi$ in addition to $x,t$. (From now on
we adopt the shorthand $\d$ for the $x$-derivative and put
$m = \hbar =1$.) Such a first-derivative term in the Schr\"odinger
equation is very natural for an equation of motion with first-order
terms in time derivatives. In the case of magnetic forces, e.g., which
are also first-order in time derivatives, such terms are present, arising
from the gauge potential. In fact, performing a $\psi$-dependent wavefunction
redefinition, we can always bring \prenew\ to the form
$$
i \dot \p = \F ( -\half \d^2 + U + V ) \f \p
= -\half \D^2 \p + ( U + V ) \p
\eqn\new$$
for some new $U$, where
$$
\f = e^{i \phi( \psi ,x,t)} \,\,
\eqn\ff$$
is now a wavefunction-dependent phase and
$$
\D = \d + i F \,\,,\,\,\,\, F = \d \phi \,\,.
\eqn\DF$$
The advantage of \new\ over \prenew\ is that it has manifestly unitary
time evolution for real $U$ and $F$.
This means, in particular, that the normalization
$$
N = \int_x \P \p
\eqn\norm$$
is preserved in time. Defining the expectation value of $x$
$$
<x> = \int_x x \P \p
\eqn\exx$$
and using \new\ we get
$$
<{\dot x}> = \int_x \P \F ( -i\d ) \f \p
= -{i \over 2} \int_x \P \D \p - ( \D \p )^* \p
\eqn\exxd$$
$$
<{\ddot x}> = \int_x \P ( -\d V -\d U + {\dot F} ) \p
\eqn\exxdd$$
In the above we integrated by parts, assuming good behavior of
$\p$ at infinity.
A sufficient condition for \new\ to have the desired classical limit
is that the equation of motion \fric\ hold at the expectation value
level. Thus we must have
$$
<{\ddot x}> = - \g <{\dot x}> - < \d V >
\eqn\must$$
and therefore
$$
\int_x \P ( -\d U + \dot F ) \p
= i{\g \over 2} \int_x \P \D \p - (\D \p )^* \p
\eqn\mmust$$
Writing $\p$ in terms of its phase and logarithmic modulus
$$
\psi = e^{\r + i\t}
\eqn\pex$$
we perform the time derivative in \mmust\ using \new. Adopting the notation
where subscripts denote partial derivative with respect to $\r$, $\t$,
$x$, $t$, we obtain after some calculation
$$
\eqalign {
i{\g \over 2} \int_x \P \D \p - (\D \p )^* \p
= \int_x & -{i \over 4} \d F_\r \left[ \P \D \p - ( \D \p )^* \p
\right] \cr
&+ {1 \over 4} F_\t \left[ \P \D^2 \p + ( \D^2 \p )^* \p -4(U+V) \P \p
\right] \cr
&+ \half \d U_\r \P \p + {i \over 2} U_\t \left[ \P \D \p - (\D \p)^* \p
\right] \cr
&+ U_\t F \P \p + (F_t - U_x ) \P \p \cr}
\eqn\long$$
The above equation must hold for arbitrary (well-behaved) $\p$.

We shall choose not to have explicit time dependence in the terms of our
equation, so that we do not spoil time invariance, and put $F_t = 0$.
We will also not allow $F$ and $U$ to be nonlocal expressions of $\p$
or functions of derivatives of $\p$, because then \new\ would become a
higher-order and/or nonlocal equation.
We notice, then, that the second term in the right hand side of
\long\ is second order in $x$-derivatives.
Therefore, since the left-hand side is first order in derivatives,
this term must vanish. Similarly, if we want the first term
to be first order in derivatives, $F_\r$ should no more contain $\p$.
So we must have
$$
F_{\r \r} = F_\t = 0 \,\,\,,\,\,\,\,{\rm or} \,\,\,\,
F = \r c (x) + c_1 (x)
\eqn\Fa$$
The $\p$-independent term $c_1 (x)$ can always be gauged away
by redefining the phase of $\p$, so we will omit it.

The relevant terms that will reproduce friction are, now,
the first and fourth terms, so we must have
$$
-{i \over 4} \d F_\r + {i \over 2} U_\t = {i \over 2} \g
\eqn\main$$
and substituting \Fa\ we obtain
$$
U = \t \Bigl( \g + \half c^\prime (x) \Bigr) + \e ( \r ,x)
\eqn\Ua$$
Finally, the remaining terms in \long\ must be a total derivative,
so that they do not contribute to the integral, that is
$$
e^{2\r} \Bigl( \half \d U_\r + U_\t F - U_x \Bigr) = \d A
\eqn\dA$$
for some $A$. Using \Ua\ and \Fa\ we get
$$
e^{2\r} \left[ \half \e_{\r \r} \d \r + \half \e_{\r x} +
(\g + \half c^\prime ) \r c - \half \t c^{\prime \prime} - \e_x \right]
= A_\r \d \r + A_\t \d \t + A_x
\eqn\dAA$$
Since there is no $\d \t$ term, $A$ must be independent of $\t$ and
thus we must have
$$
c^{\prime \prime} (x) = 0 \,\,,\,\,\,\, {\rm so} \,\,\,\,
c(x) = -2\a (x - \b )
\eqn\c$$
(the choice of factors is for later convenience). Again, $\b$ can be
shifted away with a shift in the coordinate $x$ and shall be
temporarily omitted. So $F$ becomes
$$
F = -2\a x \r
\eqn\Fb$$
For \dAA\ to be a perfect differential in $\r$ and $x$, then, we
have the consistency condition
$$
A_{\r x} = \left[e^{2\r} \half \e_{\r \r} \right]_x =
\left[ e^{2\r} \bigl( \half \e_{\r x} - (\g - \a) 2\a x \r
- \e_x \bigr) \right]_\r = A_{x \r}
\eqn\con$$
which determines the eventual form of $\e$ and $U$ as
$$
\e (\r ,x) = - \a (\g - \a ) ( \r + \half ) x^2 + \ve (\r )
$$
$$
U = (\g - \a ) \t - \a (\g - \a ) (\r + \half ) x^2 +\ve (\r )
\eqn\Ub$$
Equations \Fb\ and \Ub\ constitute the solution of \must\
and determine the Schr\"odinger equation of the particle.

The above solutions depend on two arbitrary parameters, $\a$
and $\b$ (the latter one hidden in the choice of origin),
as well as an arbitrary function of $\P \p$, $\ve (\r )$.
This last arbitrariness is present also in the ordinary case
without the friction term, since such a term does not alter the
expectation value of the equations of motion. Although its significance
is not quite clear, we shall choose to omit it, since it does not
seem to be relevant to the problem of incorporating friction.
The full-blown Schr\"odinger equation for the particle with friction,
restoring $\b$, becomes thus
$$
\eqalign{
i {\dot \p} = &-\half \Bigl( \d - i \a (x-\b ) \ln(\P \p ) \Bigr)^2 \p
+ V \p \cr
&-{i \over 2} (\g - \a ) \ln{\p \over \P} \, \p - \half \a (\g - \a )
\Bigl(1+ \ln(\P \p ) \Bigr) (x-\b )^2 \p \cr }
\eqn\Schr$$

The parameter $\a$ is interesting. Apparently, it can take any real
value. Notice, however, that for any normalizable wavefunction,
$\p$ must vanish fast enough at $x \to \pm \infty$ and thus $\ln{\P \p}
\to -\infty$. Therefore, the second term in the potential $U$ will be
badly behaved if $\a (\g - \a ) < 0$ and the above equation will
be unstable. To avoid this, we must have
$$
0 \leq \a \leq \g
\eqn\ineq$$
Therefore, the range of $\a$ is restricted by the magnitude of
$\g$. In particular, for $\g = 0$ this parameter is absent, which
shows that it is specific to the friction problem.
Also, the parameter $\b$ becomes irrelevant when $\a = 0$ and
thus when $\g = 0$, which shows that it also is a characteristic
of the friction.
For the extreme value $\a = 0$ we recover Kostin's equation.
For the other extreme value $\a = \g$ we obtain an equation
without additional potentials but with a ``magnetic" (gauge)
term.

Indeed, we can view $F$ as the spatial component of
a gauge field and $U$ as the time component of the same field.
Notice, then, that the quantity ${\dot F} - \d U$ which appears
in \mmust\ is the field strength of the above gauge potential.
One could worry, then, that since the above equations for all
values of $\a$ reproduce the same equation \mmust, they are
gauge equivalent. This, however, is not the case. Although it
is true that all equations have the same {\it expectation value}
of the field strength (the integral in \mmust), they differ
locally by total derivative terms. Therefore, although they all
have the same classical limit, they describe different
quantum mechanics. In fact, if we wanted to perform a gauge
transformation so as to gauge away the spatial component $F$ and
trade it for a potential term, then we would end up with a
potential which would contain nonlocal terms in the wavefunction.
This nonlocality is, however, an artifact of the temporal gauge $F=0$,
since in our gauge \Schr\ is perfectly local. This nonlocality is,
perhaps, the reason why our equation was not previously discovered
in Kostin's approach.

A remarkable property of \Schr\ is that, in spite of its nonlinearity,
it is still invariant under rescaling of $\p$. This means, in particular,
that physics does not depend on the normalization of the wavefunction.
Specifically, the transformation
$$
\p \to \exp \left[ i\t_o e^{-(\g - \a ) t} + i\a \r_o (x-\b)^2 + \r_o
\right] \p
\eqn\trans$$
with $t_o$, $\r_o$ constants, leaves \Schr\ invariant and constitutes
the generalization of the original complex rescaling transformation
of the linear Schr\"odinger equation for the case of nonzero friction.
Notice that, for $\a \neq \g$, it explicitly involves time and for $\a
\neq 0$ it explicitly involves the space coordinate.

The original Kostin's equation exhibited stationary nondissipating
states. In fact, it is easy to see that {\it all} the energy
eigenstates of the hamiltonian without friction remain stationary
states for Kostin's equation. The dissipation, then, appears only upon
mixing these states due to the nonlinear nature of the equation. It is also
easy to see that the special case $\a = \g$ of \Schr\ also exhibits the
same stationary states. It is less clear, however, whether \Schr\ for
arbitrary $\a$ will have such states. A stationary state has the form
$$
\p = e^{i \t (x,t)} \chi (x) \,\,.
\eqn\sta$$
Therefore, the gauge potential term $F$ becomes now independent
of time and can be gauged away through the redefinition
$$
\t = \varphi + \int_{y= x_o}^x \a y \ln ( \chi^* \chi )
\eqn\redef$$
where $x_o$ is an irrelevant constant (we have shifted $\b$ back to zero
for convenience). Using, now, \Schr\ we find that the phase $\varphi$ will
be independent of $x$ while $\chi$ will satisfy the nonlinear eigenvalue
equation
$$
\left[ -\half \d^2 + V + \a (\g - \a ) \left\{ \int_{y= x_o}^x y \ln
( \chi^* \chi ) - \half \Bigl(1 + \ln ( \chi^* \chi ) \Bigr) x^2 \right\}
-{i \over 2} (\g - \a ) \ln {\chi \over \chi^*} \right] \chi = E \chi
\eqn\eigen$$
with the phase satisfying
$$
- {\dot \varphi} = E + (\g - \a ) \varphi \,\,
\eqn\phase$$
The eigenvalue $E$ in general changes upon rescaling of $\chi$ and
thus does not correspond to the energy of the particle. Also, although
in the frictionless case $\chi$ can always be chosen real, this is
not necessarily true any more in the present nonlinear case.

It is clear that for $\a =$ 0 or $\g$ \eigen\ is identical to the
eigenvalue problem of the frictionless hamiltonian and we recover
all energy eigenstates as stationary states. For arbitrary $\a$
we cannot, at the moment, make a general statement on the existence
of such states, although it is clear that if they exist they will be
different from the ordinary ones. This is physically appealing, since
we expect friction to alter the properties of the states of the system.
In fact, the value of $\a$ may be fixed by further examining this issue.

We can examine some qualitative properties of \eigen. Notice that,
if the wavefunction $\chi$ is real and falls off exponentially at
infinity with some power law in $x$, that is,
$$
\ln \chi \to - \kappa |x|^n ~~{\rm for}~~ |x| \to \infty
\,\,,\,\,\,\,\kappa ,n > 0
\eqn\power$$
then the $\chi$-dependent potentials in \eigen\ will behave as
$$
{\a (\g - \a ) \kappa n \over n+2} |x|^{n+2}
\eqn\beh$$
confirming that the above equation is stable for $0 < \a < \g$
in this case. It is tempting to speculate that the physically relevant
value of $\a$ is the one achieving maximal stability, that is,
$\a = \half \g$. To further check this, we will examine
the case of an external harmonic potential at $x_o$:
$$
V(x) = \half \omega^2 (x - x_o )^2
\eqn\harm$$
and treat the friction perturbatively, that is, assume $\g << \omega$.
Starting with the harmonic oscillator ground state as the zeroeth-order
wavefunction and plugging it into the left-hand side of \eigen\ we
obtain a first-order effective potential
$$
V_1 (x) = \half {\tilde \omega}^2 ( x - {\tilde x}_o )^2
+ {1 \over 4} \a (\g - \a ) \omega x^4 + {\rm constant}
\eqn\VV$$
where
$$
{\tilde \omega}^2 = \omega^2 - \a (\g - \a ) \,\,,\,\,\,\,
{\tilde x}_o = {\omega^2 \over {\tilde \omega}^2} x_o \,\,.
\eqn\til$$
We see that the first-order effect of friction is a shift of the origin
of the harmonic potential (of which we will say more later) as well as
a change of its strength, in addition to the appearence of anharmonic
terms. If we choose now $\a = \half \g$ we get
$$
{\tilde \omega}^2 = \omega^2 - \left({\g \over 2}\right)^2 \,\,.
\eqn\newo$$
This is exactly the shift in the frequency of damped oscillations of
an oscillator with friction, further corroborating this choice. The
anharmonic terms account for the quantum mechanical effects of the
friction and ensure the stability of the problem even for the
overdamped case $\g > 2 \omega$ (in this case, of course, perturbation
theory is not valid).

Remarkably, when the external potential is zero we can find an {\it exact}
stationary state for the problem. It is easy to check that
$$
\chi = C \exp\left[ -{\a ( \g - \a ) \over 12} x^4 \right]
\eqn\stat$$
with $C$ a constant, is a solution of \eigen. Thus, for $\a \neq 0,\g$,
the particle has a localized stationary state in the absence of external
potentials, due entirely to friction. Even more remarkably, the particle
exhibits spontaneous violation of translation invariance. Remember that
the origin $x=0$ is fixed by the requirement that the term $\b$ in
\c\ vanish. The wavefunction \stat\ then is centered around this
special point. Choosing a different value for $\b$ would place
this state at the point $x = \b$. The parameter $\b$ then is
a symmetry breaking parameter for the problem, due entirely to friction
(remember that $\b$ becomes irrelevant when $\g = 0$).

A similar effect is also evident in the first-order potential \VV\ in
the problem with the harmonic potential, manifesting in the shift of the
origin of the harmonic forces. If this point coincided with the point
$x=\b$, there would be no such shift.
It is not clear, of course, that this effect would survive higher-order
in $\g$ effects. At the classical limit, for instance, such a shift
cannot occur and thus it must be washed out by nonlinear effects.

It seems odd that there should be spontaneous symmetry breaking in
a finite degrees of freedom problem. It is not so surprising, however,
in the picture where friction is reproduced by coupling the particle
to a continuous infinity of harmonic oscillator degrees of freedom,
whose frequency extends all the way down to zero. It is conceivable
that an appropriate choice of such a system exhibits spontaneous
symmetry breaking. Clearly this point deserves further investigation.
We find it curious, however, that our direct approach to the problem
should invoke such ``memories" of a possible infinite dimensional
description.

A property of Kostin's equation was that it involved the phase $\t$
of the wavefunction which is defined only modulo $2\pi$. This is
inconsequential in most cases, since this only amounts to a
time-dependent redefinition of the phase of the solutions (see \trans).
It is problematic, however, on spaces with topologically nontrivial
loops, since, in general, there is no way of defining a single-valued
phase on such spaces. For instance, Kostin's equation cannot deal
with a particle on the circle. (This is related to the fact that a
smooth decay of the energy is incompatible with momentum quantization
on the circle.) Our equation with $\a = \g$, on the other
hand, does not involve the phase. It exhibits nevertheless a similar
problem since it is not explicitly translation invariant, due to the
gauge field $F$. (This is related to the parameter $\b$, as explained
above.) It is not clear, then, how it can manifest the correct
periodicity on the circle. The case of general $\a$ presents a mixture
of the two problems.

Finally, we point out that the nonlinearity of the above equations is
not unfamiliar. Viewed as field equations, it simply means that
they represent interacting theories which, in the second quantized picture,
have particle creation and annihilation. This is very much in line
of the description of friction as an interaction process with a
reservoir of other particles. It should be noted, however, that there
is no action which gives \Schr\ as a lagrangian equation of motion
for any $\a$. The non-hamiltonian nature of the problem emerges,
then, in the next level of quantization. Therefore, the second
quantization of the above theories is an open issue.

\ack{
A.P.P. was supported in part by D.O.E. grant DE-AC02-76ER-02271;
R.T. thanks the Spanish Ministry of Education for a Fellowship.}

\refout
\end